\documentclass[trackchanges]{aastex701}
\usepackage{comment}

\usepackage{showyourwork}

\begin{document}

\title{Non-detection of J1407~b in ALMA Band 7 observations}

\author[orcid=0000-0001-9443-0463,sname='Klaassen']{Pamela Klaassen}
\affiliation{UK Astronomy Technology Centre, Royal Observatory Edinburgh, Edinburgh, EH9 3HJ, UK}
\email[show]{pamela.klaassen@stfc.ac.uk}  

\author[orcid=0000-0002-7064-8270,sname='Kenworthy']{Matthew A. Kenworthy} 
\affiliation{Leiden Observatory, Leiden University, Postbus 9513, 2300 RA Leiden, The Netherlands}
\email[show]{kenworthy@strw.leidenuniv.nl}

\author[0000-0003-2008-1488]{Eric E. Mamajek}
\affiliation{Jet Propulsion Laboratory, California Institute of Technology, Pasadena, CA 91109, USA}
\email{mamajek@jpl.nasa.gov}

\author[orcid=0000-0003-2458-9756,sname='Marel']{Nienke van der Marel}
\affiliation{Leiden Observatory, Leiden University, P.O. Box 9513, 2300 RA Leiden,The Netherlands}
\email{nmarel@strw.leidenuniv.nl}  

\author[orcid=0000-0001-5778-0376,sname=‘Min’]{Michiel Min}
\affiliation{SRON Space Research Organisation Netherlands, Leiden, The Netherlands}
\email[show]{M.Min@sron.nl}  

\author[orcid=0000-0002-5510-8751,sname=’triaud']{Amaury H.M.J. Triaud}
\affiliation{School of Physics \& Astronomy, University of Birmingham, Edgbaston, Birmingham B15 2TT, United Kingdom}
\email{a.triaud@bham.ac.uk}  

\author[0000-0001-5073-2849]{Antonio S. Hales}
\affiliation{National Radio Astronomy Observatory, 520 Edgemont Road, Charlottesville, VA 22903-2475, United States of America}
\affiliation{Joint ALMA Observatory, Avenida Alonso de Córdova 3107, Vitacura 7630355, Santiago, Chile}
\affiliation{Millennium Nucleus on Young Exoplanets and their Moons (YEMS), Chile}
\email{antonio.hales@alma.cl}

\begin{abstract}

We report on ALMA Band 7 continuum observations towards the star 1SWASP J140747.93-394542.6 taken in mid 2024. These observations were a follow-up of a previous detection of an emission source in the J1407 field of view at an unexpected position in 2017. Proper motion analysis indicated that if this were the object responsible for the 2007 eclipse of J1407, it would be detectable at a new position, but still within an ALMA field of view in 2024.
Here we present the non-detection of emission in the ALMA field of view (to a 1$\sigma$ upper limit of 17.5 $\mu$Jy),at both the 2017 position, and the position expected from proper motions. We place upper limits on a source at the proper motion corrected expected location in this 2024 data, and rule out the possibility of a dusty object being responsible for the 2007 eclipse.

\end{abstract}


\section{Introduction}

A complex series of dimmings of the young star 1SWASP J140747.93-394542.6 (hereafter shortened to J1407) in May 2007 was hypothesised to be caused by a giant circumsecondary disk transiting the disk of the star \citep{Mamajek13,Kenworthy15}.
No other eclipses have been seen to date towards J1407, and none are present in archival data \citep{Mentel18}.
Initial ALMA observations were undertaken in 2017 to determine whether the dust emission from the eclipsing object could be detected in Band 7 (345 GHz) continuum emission. With an rms noise level of 18.6$\mu$Jy, an unresolved source was seen in the ALMA data \citep{Kenworthy20} with a peak flux of 79$\mu$Jy, spatially offset from the position expected for J1407 given its proper motion, suggesting a chance transit. The distance between the 2007 position of J1407 and the 2017 flux detection suggested a proper motion consistent with the ALMA detected object having moved sufficiently to be detectable at a new, spatially separate, position in 2024. Here we present a second epoch follow-up of those original ALMA observations to attempt to detect that object at its new position.

\section{Observations and Data Processing} 

We observed J1407 in continuum emission in Band 7 using ALMA between 26 June and 13 July 2024 for project 2023.1.001488.S (PI: M. Kenworthy). The array was in configuration C-8 and observations were taken in TDM mode, giving high resolution continuum observations. J1407-4302 was used as both bandpass and phase calibrator and J1427-4206 was used for amplitude calibration. The data were pipeline processed, and the individual measurement sets were concatenated in the uv plane before imaging. Imaging was peformed with a manual mask and allowed to progress for 1000 iterations before stopping. The final image has a synthesized beam of $0.12''\times0.07''$ at pa=111$^\circ$, and an rms noise limit of 17.5$\mu$Jy, close to that of the previous observations.

Figure \ref{fig:J1407_epoch_comparison} shows both epochs of  ALMA observations of the region: from 2017 (left) and 2024 (right). Also labelled on these plots are: (1) the location of J1407 when the eclipse occured in 2007, (2) the position of the $\sim4\sigma$ ALMA detection in 2017 and the expected position of J1407 in the same epoch, as well as (3) the projected positions of the ALMA source and J1407 in 2024. The third epoch (2024) positions of the ALMA source and J1407 were derived using the built in proper motion functionality in \texttt{astropy.coordinates} having derived the proper motion from the first and second epoch observations in (1) and (2) above.
%






\section{Results and Discussion}

\begin{figure*}
    \script{plot_proper_motion_figure.py}
    \includegraphics[width=1\textwidth]{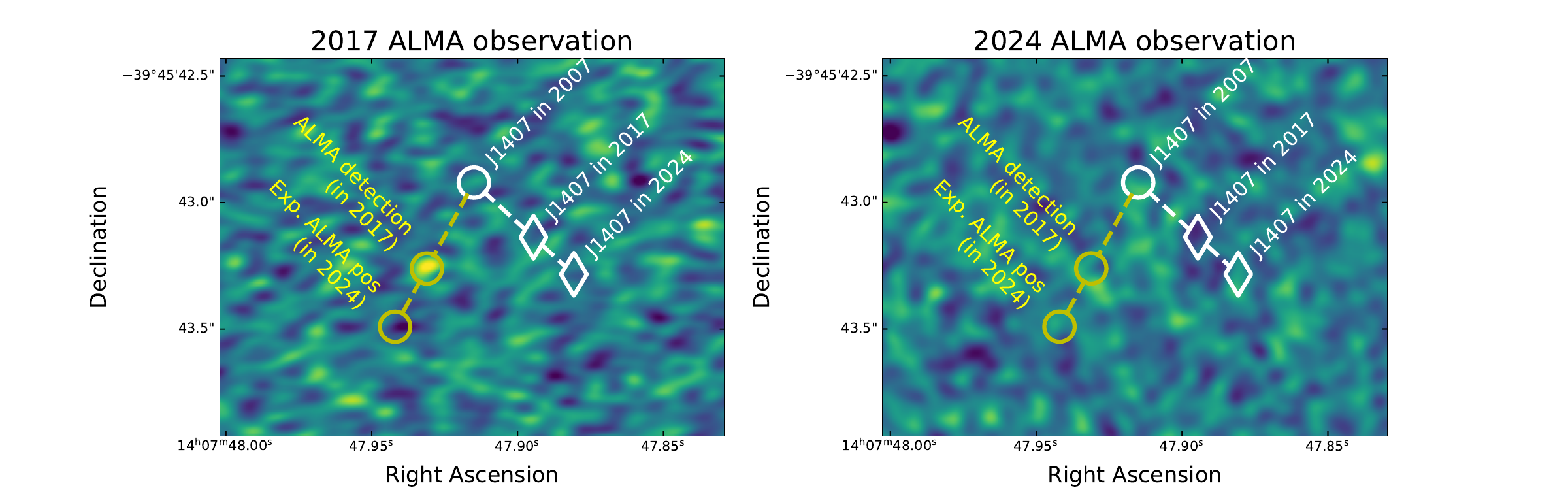}
    \caption{Multi-epoch ALMA observations of the J1407 area. There was a $\sim4\sigma$ 345 GHz continuum detection in the 2017 epoch observations that does not seem to persist to the 2024 observations, despite the two observations having similar beams and noise levels.}
    \label{fig:J1407_epoch_comparison}
    
\end{figure*}

As can be seen in Figure \ref{fig:J1407_epoch_comparison}, there is no detection of 345 GHz continuum emission at the projected location of the 2017 ALMA continuum source had it been responsible for the 2007 eclipse of J1407. If the target continued to be as bright as in 2017, it should have been detected at the $\sim4\sigma$ level, but there is nothing above the rms noise level in the emission map at this position.

There were a few models put forward in \cite{Kenworthy20} for what could be causing that emission source in the 2017 ALMA data: a free-floating ring system that had a chance alignment with J1407 in 2007 or an unrelated extragalactic background object unrelated to the eclipse. The latter was dismissed based on statistically small chances of an object with that flux being detected that close to J1407. The former was deemed more realistic because the proper motion of that emission source on that 10 year baseline was consistent with the derived eclipsing objects transverse velocity calculated in \cite{Kenworthy15}. 

Comparing our current noise levels to the dust models presented in \cite{Kenworthy20}, thick disks with dust grains larger than 1mm, as favoured in the previous paper, \textit{of any mass} are now ruled out.  This suggests that the original detection was likely a noise spike, or some form of time varying or dimming object. Had it been a persistent background source, it should have remained detectable at the $\sim4\sigma$ level at the same position as the 2017 observation.

\facilities{ALMA}

\begin{acknowledgements}

This paper makes use of the following ALMA data: ADS/JAO.ALMA\#2023.1.01488.S. ALMA is a partnership of ESO (representing its member states), NSF (USA) and NINS (Japan), together with NRC (Canada), MOST and ASIAA (Taiwan), and KASI (Republic of Korea), in cooperation with the Republic of Chile. The Joint ALMA Observatory is operated by ESO, AUI/NRAO and NAOJ. Part of this research was carried out at the Jet Propulsion Laboratory, California Institute of Technology, under a contract with the National Aeronautics and Space Administration (80NM0018D0004).

This study was carried out using the reproducibility software \href{https://github.com/showyourwork/showyourwork}{\showyourwork}
   \citep{Luger2021}, which leverages continuous integration to programmatically download the data from \href{https://zenodo.org/}{zenodo.org}, create the figures, and compile the manuscript. Each figure caption contains two links: one to the dataset stored on zenodo used in the corresponding figure, and the other to the script used to make the figure (at the commit corresponding to the current build of the manuscript). The git repository associated to this study is publicly available at \url{https://github.com/mkenworthy/J1407b_alma_2024}. The datasets are stored at \url{https://doi.org/10.5281/zenodo.17670200}.

\end{acknowledgements}

\software{astropy \citep{astropy:2013,astropy:2018,astropy:2022}
        casa \citep{CASA2022}
        showyourwork \citep{Luger2021}
}

\bibliography{kenworthy}{}
\bibliographystyle{aasjournalv7}

\end{document}